\documentstyle[aps,prl,multicol,epsfig,amssymb]{revtex}
\begin{document}
\def\be{\begin{equation}}
\def\ee{\end{equation}}
\def\bea{\begin{eqnarray}}
\def\eea{\end{eqnarray}}
\def\E{{\rm e}}
\def\bearst{\begin{eqnarray*}}
\def\eearst{\end{eqnarray*}}
\def\peleven{\parbox{11cm}}
\def\peffec{\peight{\bearst\eearst}\hfill\peleven}
\def\pspace{\peight{\bearst\eearst}\hfill}
\def\ptwelve{\parbox{12cm}}
\def\peight{\parbox{8mm}}
%\twocolumn[\hsize\textwidth\columnwidth\hsize\csname@twocolumnfalse\endcsname

\title{Closure of two dimensional turbulence: the role of pressure gradients}
 \author{G.~Boffetta$^{1}$, M.~Cencini$^{2,3}$ and J.~Davoudi$^{4,5}$}
\address{$^{1}$ Dipartimento di Fisica Generale and INFM,
Universit\`a di Torino, 
Via Pietro Giuria 1, I-10125 Torino, Italy}
\address{$^{2}$ Dipartimento di Fisica, and INFM,
Universit\`a di  Roma "la Sapienza", 
Piazzale Aldo Moro 5, 00185 Roma, Italy}
\address{$^{3}$ Observatoire de la C\^ote d'Azur - Nice
Boulevard de l'Observatoire, B.P. 4229 F-06304 NICE Cedex 4}
\address{$^{4}$ Max-Planck-Institute f\"ur Physik Komplexer Systeme, 
N\"othnitzer Str. 38, 01187 Dresden, Germany}
\address{$^{5}$ Fachbereich Physik, Philipps Universit\"at, Renthof 6,
35032 Marburg, Germany}            

\maketitle
\begin{abstract}
Inverse energy cascade regime of two dimensional turbulence is
investigated by means of high resolution numerical simulations.
Numerical computations of conditional averages of transverse pressure
gradient increments are found to be compatible with a recently proposed
self-consistent Gaussian model.
An analogous low order closure model for the longitudinal pressure gradient
is proposed and its validity is numerically examined. In this
case numerical evidence for the presence
of higher order terms in the closure is found.
The fundamental role of conditional statistics between longitudinal
and transverse components is highlighted.
\\ PACS number(s)\,:
47.27.-i, 47.27.Ak, 05.40.-a
\end{abstract}
\begin{figure} 
\vspace{-.3truecm}
\centerline{\includegraphics[scale=0.60,draft=false]{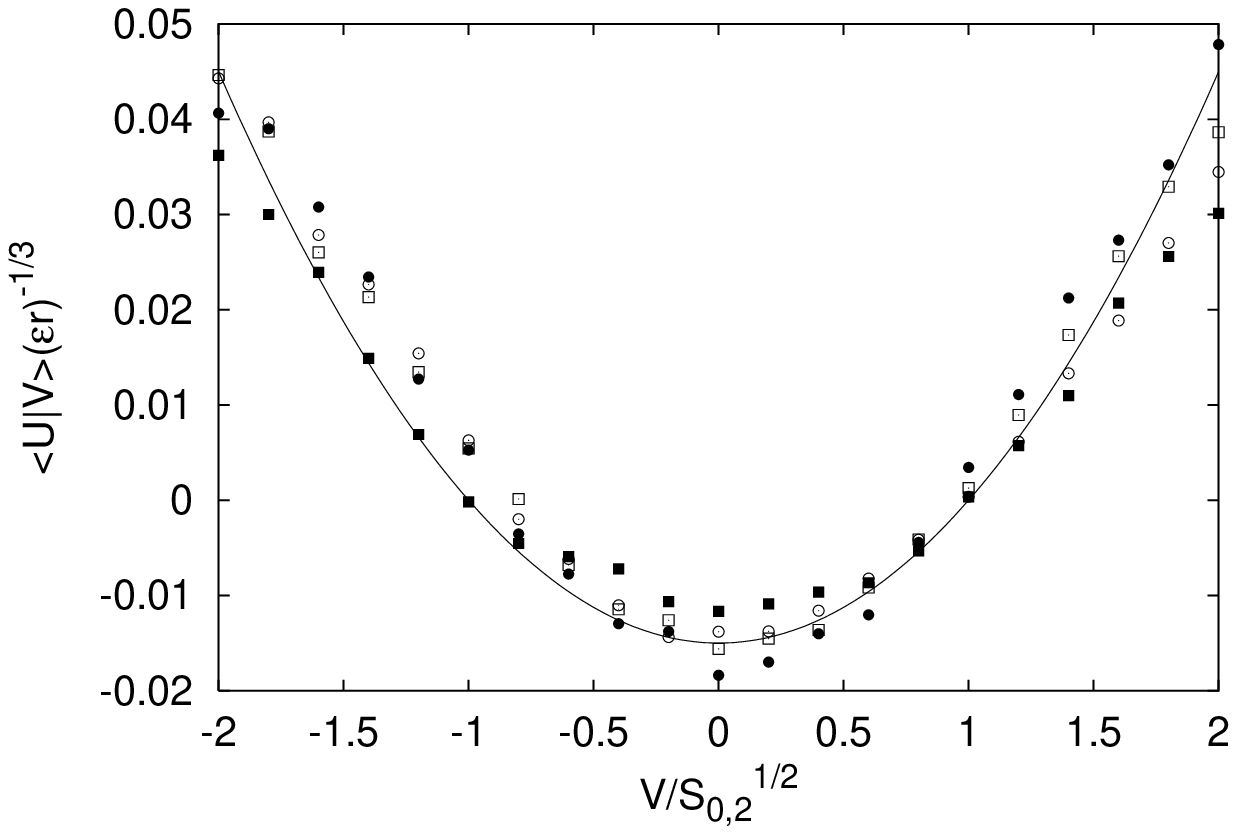}}
\centerline{\includegraphics[scale=0.60,draft=false]{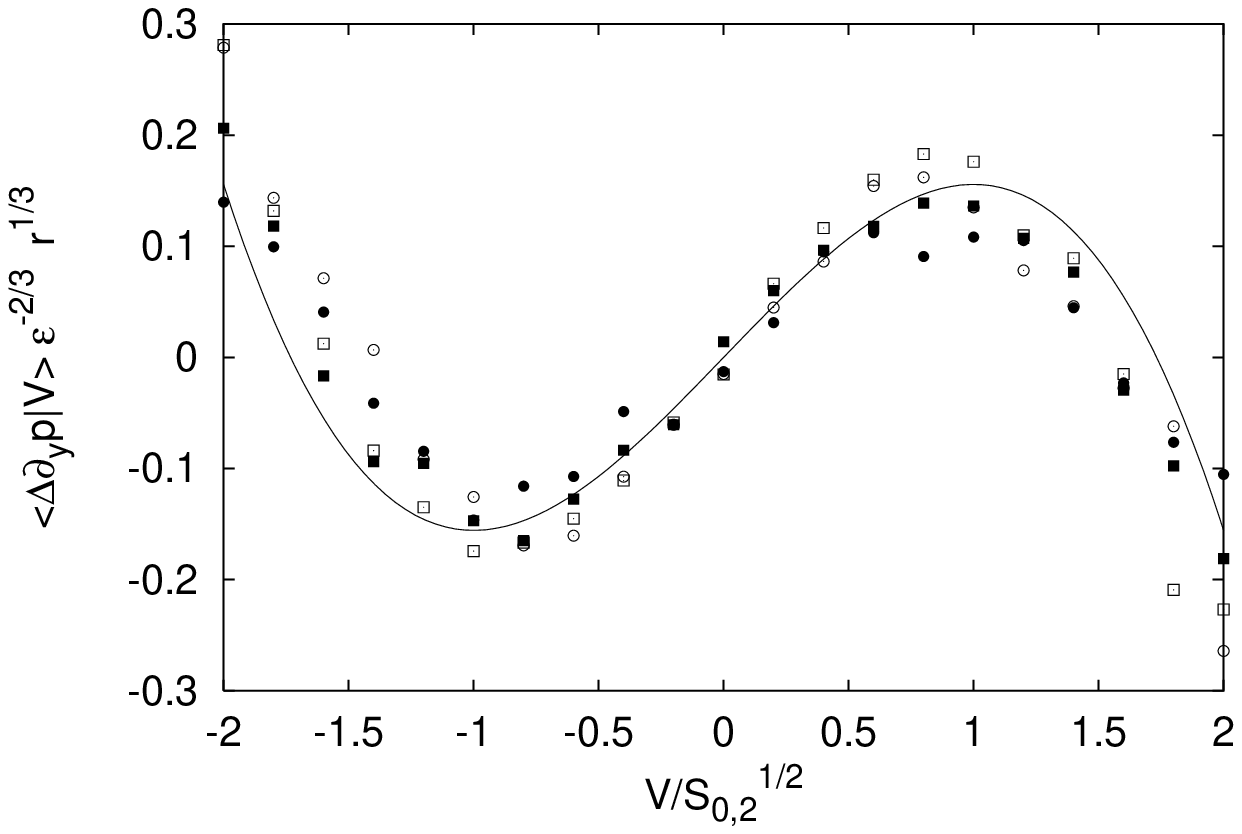}}
\vspace{0.4truecm}
\narrowtext
\vspace{-.2truecm}
\caption{(a) $\langle U|V,r\rangle$ and (b) $\langle \Delta \partial_y
p|V,r \rangle$ computed at $r= 0.025$ (boxes) and $r= 0.037$ 
(circles).  Empty symbols refer to the Gaussian forcing and
full ones to the restricted in wavenumber one.  The full lines
represent predictions (\ref{pred1}) and (\ref{pred2}) with
$A_{2,0}=11$. }
\label{fig1}
%\vspace{-.3truecm}
\end{figure}

%\pacs{PACS number(s)\,: }
%%%%%%%%%%%%%%%%%%%%%%%
%
\begin{multicols}{2}
\noindent 
The existence of two simultaneous inertial ranges in two-dimensional
turbulence, as a consequence of coupled energy and enstrophy
conservation, is one of the most important phenomena in statistical
fluid mechanics \cite{kraichnan_2d}.  At variance with
$3D$-turbulence, the energy injected into the system at scale $\ell_f$
flows toward the large scales, while the enstrophy cascades down on
the small scales. 
Because of the inverse energy cascade, the
Navier-Stokes equations,
\begin{equation}
\partial_t u_i + u_j \partial_j u_i = - \partial_i p + \nu \partial^2
u_i+ f_i
\label{eq:1}
\end{equation}
which rule the evolution of an incompressible ($\partial_i u_i= 0$)
velocity field, cannot reach a steady-state unless an energy sink at
large scales is added.  Alternatively one can consider an ensemble of
solutions of (\ref{eq:1}) with a fixed energy value below the
condensation level~\cite{SY93}, i.e. with an integral scale $L(t)$
(growing in time as $t^{3/2}$) still much smaller than the system
size. Because of the scaling of the characteristic times, the small
scales (inertial range) in the system $\ell_f\ll r \ll L$ can be
considered in a
stationary state. One of the most challenging problems is to
understand the statistics of velocity fluctuations $\Delta {\bf
u}({\bf r})={\bf u}({\bf x}+{\bf r})-{\bf u}({\bf x})$ \cite{F95}.
In homogeneous and isotropic turbulence it amounts to study the {\it
joint} probability density function (PDF) $P(U,V,r)$ of longitudinal
$U$ and transverse $V$ velocity differences where
$\Delta {\bf u}=U {\bf \hat{x}}+V{\bf \hat{y}}$ and
${\bf \hat{x}}=\frac{{\bf r}}{r}$.
Recently experimental \cite{PT98} and numerical \cite{BCV00}
investigations
in two dimensions have shown that the probability distribution of the
{\it pure} longitudinal $P(U,r)$ and transversal $P(V,r)$ velocity
differences at inertial scales 
display a close-to-Gaussian statistics with undetectable intermittency
corrections to structure function exponents. 
Although the establishment of normal 
scaling in all inverse cascades seem
to be generic \cite{fal} nevertheless the Gaussianity of the statistics in
inverse cascade of the forced two dimensional turbulence is remained
to be understood.  
From
(\ref{eq:1}), a set of equations for generic mixed structure
functions, i.e. $S_{n,m}(r)\equiv\langle
U^nV^m\rangle= A_{n,m} r^{\xi_{n,m}}$ have been obtained
\cite{Hill,Y01}.  In \cite{Y01} those equations are elaborated from
the joint PDF equation.  Unfortunately, the PDF equation is not
closed, resembling the well known closure problem in
turbulence. 
In the inverse energy cascade regime,
dissipative contributions
can be neglected so the remaining unclosed terms are the
longitudinal and transversal pressure gradients increments. 
Recently Yakhot \cite{Y01,Y99} suggested a
self-consistent model for the pressure gradient
increments and succeeded to obtain a Gaussian distribution for the
transverse PDF, $P(V,r)$. 
%The Gaussian 
%result of the proposed effective model is consistent with experiments
%\cite{PT98} and numerics \cite{BCV00}. 
Although the experimental \cite{PT98} and numerical \cite{BCV00}
observations support the Gaussian result of the effective low order
model, nevertheless a direct numerical computation of
the pressure gradient increment contribution is
still lacking. \\ 
The main aim of this work is to compare the numerical evaluation \cite{simul}
of transverse and longitudinal components of pressure gradient
increments with the theoretical predictions of a recently introduced
closure scheme. For the first time, we emphasize the importance 
of velocity mixed conditional averages such as
$\langle U|V,r \rangle$ and 
$\langle V^2|U,r\rangle$ generally arising
in the pure longitudinal or transversal PDF equations.
To our surprise the existence of such objects has been 
neglected in all the previous theoretical modelings.   
As an essential step for the description of pure velocity statistics
we numerically evaluate the behavior of these new objects for which 
some effective models are proposed. 
Such an investigation provides a direct check of the closure model. 
%For what concerns the transverse pressure gradient increment, the second
%order closure 
%proposed in \cite{Y01} together with conditional velocity
%average $\langle U|V,r \rangle$ are in remarkable agreement with the data
%within two standard deviations of $V$ statistics.
%Also a second order closure for the longitudinal pressure gradient
%increment and a first order one for conditional
%velocity average $\langle V^2|U,r\rangle$ are approximately
%consistent with their direct numerical evaluation in two standard
%deviation of $U$ statistics. Nevertheless the numerical
%values of the hyper-skewness, which are related to dynamical features
%of the cascade, directly evaluated from the
%data, are inconsistent with the predicted values calculated from the low 
%order closure. \\
\\
By standard statistical tools \cite{Y01,MY}, starting from the
Navier-Stokes
equations (\ref{eq:1}), it is possible to derive the following exact
PDF equation for joint transversal and longitudinal velocity
increments:
\begin{eqnarray}
\label{pdf-joint}
&&\left[{\partial}_r U+{U \over r}-{1 \over r}{\partial}_V U
V+{1 \over r} {\partial}_U V^2 \right] P(U,V,r)=\nonumber \\
&&\left[\varepsilon ({\partial}_U^2+{\partial}_V^2)
+{\partial}_U  {\cal P}_{x,u}
+{\partial}_V {\cal P}_{y,v}\right] P(U,V,r) \,, 
\end{eqnarray}
where $\varepsilon\equiv \langle f_i u_i\rangle$ is the rate of energy input and the conditional
transversal, ${\cal P}_{y,v}\equiv\langle \Delta \partial_y p|U,V,r
\rangle$, and longitudinal, ${\cal P}_{x,u}\equiv\langle \Delta
\partial_x p|U,V,r \rangle$, pressure gradient increments are the only
unclosed terms. In pure longitudinal and transversal PDF equations
other unknown quantities play role. Indeed, by integrating
(\ref{pdf-joint}) over $U$ or $V$ the terms 
$\langle U|V,r\rangle=\int_{-\infty}^{+\infty} U P(U|V,r) dU$ and
$\langle V^2|U,r\rangle=\int_{-\infty}^{+\infty} V^2 P(V|U,r) dV$ 
appear in the pure transverse or longitudinal PDF
equations respectively, pinpointing the statistical
dependence between longitudinal and transversal components.
Let us start with the transversal one, for which the knowledge of
$\langle \Delta \partial_y p | V,r\rangle=
\int_{-\infty}^{+\infty}{\cal P}_{y,v}P(U|V,r) dU$ and of 
$\langle U|V,r\rangle$ is sufficient to close the equation. Following the
recently proposed closure \cite{Y01}, we {\it assume} a second order
expansion for the transverse pressure gradient increments ${\cal
P}_{y,v}$ in terms of {\it local} velocity increments $U$ and $V$.
Even if the locality assumption is not based on rigorous grounds, 
there are some arguments in support of its plausibility \cite{Y99}.
Once locality is accepted, keeping only second
order terms is motivated from the fact that for Gaussian fields only
quadratic combinations of $U$ and $V$ appear~\cite{gotoh1}. Some
physical constraints simplify the expansion even
further~\cite{nota,nota2}, ending with Yakhot ansatz \cite{Y01}
\begin{equation}
\label{pre-eq1} 
\langle \Delta \partial_y p|U,V,r \rangle = -h
\frac{U V}{r} - b(\varepsilon r)^{1/3}\frac{V}{r} \,.  
\end{equation}
To directly check the closure one has to compute quantities like
${\cal P}_{y,v}$.  However, to be more quantitative, here we
numerically compute $\langle \Delta \partial_y p|V,r
\rangle$ and $\langle \Delta \partial_y p|U,r \rangle$ for which we
have a better statistics.  For symmetry reasons $\langle \Delta
\partial_y p|U,r\rangle\!=\!0$ as confirmed by simulations, and we are
left with the analysis of the term $\langle \Delta \partial_y p|V,r
\rangle$.  We start by writing the quantities of interest in a scale
invariant form. For a scale invariant solution for the PDF equation,
i.e. $P(V,r)\!=\!P(V/(\varepsilon r)^{1/3})\equiv P(X)$, is sufficient
to require scale invariance of $\langle U|V,r\rangle$ and $\langle
\Delta
\partial_y p|V,r\rangle$. We thus define $\langle U|V,r\rangle\!
=\!(\varepsilon r)^{1/3}F(X)$ and $\langle \Delta \partial_y
p|V,r\rangle\!=\!  [(\varepsilon r)^{2/3}/r]G(X)$. 
The major
challenge now is to determine the functional form of $G(X)$ and $F(X)$.
Taking into account the symmetries of (\ref{eq:1}), we assume for $\langle
U|V,r\rangle$ an even polynomial expansion in $V$.  Invoking the
homogeneity, $\overline{\langle U | V,r \rangle}\!=\!0$, leads to the
low order expansion:
\begin{equation}
\label{f-eq}
F(X)=C_2(-A_{0,2}+X^2)\,,
\end{equation}
meaning that positive (negative) longitudinal velocity increments
correspond to large (small) transverse velocity increments.
Furthermore, by integrating (\ref{pre-eq1}) over $U$ one obtains $G(X)=
- h X F(X) - b X$.  Apparently this is a two parameter expansion,
however the constraint $\overline {V {\cal P}_{y,v}}\!=\!0$
\cite{Hill,Y01} implies $h \overline {X^2 F(X)} \!=\!-b\overline
{X^2}$. Since $\overline {X^2 F(X)}\!=\!  A_{1,2}\!=\!1/2$, one ends
up with the relation $h A_{1,2}\!=\!-b A_{0,2}$.  The important fact
is that this expansion is consistent with Gaussianity of transverse
fluctuations and also gives a reasonable account for pressure
contributions in the structure function equations \cite{Hill,KS01,BCD01}.
Indeed plugging the expansion for $F$ and $G$ in the dimensionless
transverse PDF equation, one obtains the Gaussian result
$P(X)=\exp(-X^2/{2A_{0,2}})$ \cite{Y01,BCD01}, which is consistent with
simulations and experiments \cite{PT98,BCV00}.  Since positivity and
finiteness of the PDF fixes the constant $C_2\!=\!1/(4 A_{0,2}^2)$ and
$h=4/3$ therefore $A_{2,0}=3/5 A_{0,2}$ is the only free parameter of
the theory \cite{Y01,BCD01}.
Therefore, within second order approximation one has 
\begin{eqnarray}
\label{pred1} 
&&\langle U|V,r\rangle ={ \footnotesize \frac{(\varepsilon r)^{1/3}}{4
A_{0,2}^2}}\left(-A_{0,2}+X^2\right)\,,\\ 
&&\langle \Delta \partial_y p|V,r
\rangle = {\footnotesize \frac{(\varepsilon r)^{2/3}}{r}} \left(
{\footnotesize \frac{X}{A_{0,2}}-\frac{X^3}{3A_{0,2}^2}} \right)\,,
\label{pred2}
\end{eqnarray} 
which up to about two standard deviations agree remarkably well with
the numerical data (see Fig.~\ref{fig1}). Moreover, using
(\ref{pred1}) as a fitting function, we obtained $A_{2,0} = 11 \pm 1$ 
which is close, within the statistical errors, with
the value obtained in previous experimental \cite{PT98,sommeria},
numerical \cite{SY93,BCV00,Vallis}) and analytical \cite{Olla}
studies. We remark that the good agreement of DNS data
with (\ref{pred2}) provides a first evidence (even
if numeric) for the plausibility of the locality assumption.  
%%%%%%%%%%%%%%%%%%%%%%%%
%%%%%%%%%%%%%%%%%%%%%%%%%%%%%%%%%%%%%%%%%%%%%%%%%  
\noindent 
However one can verify that assuming higher order polynomials for $F(X)$, 
can result in non zero higher order terms in $G(X)$.  
Indeed plugging the Gaussian result in
the equation for $P(X)$, for any order consistent with Gaussianity, $G(X)$
is
expressible as a functional of $F(X)$. So 
we obtain \cite{BCD01},
\bea 
G(X)=&&-\frac{X}{A_{0,2}}-\frac{4}{3}\bigg( XF(X) \nonumber \\
&+&e^{\frac{X^2}{2A_{0,2}}}\int^X F(X') e^{-\frac{X'^2}{2A_{0,2}}}
dX'\bigg)
\label{highorders}
\eea Plugging the self consistent low order model (\ref{f-eq}) in
(\ref{highorders}) will reduce the proposal of Yakhot. Equation
(\ref{highorders}) provides a way to generalize the model $G(X)$, 
in a self consistent way, to
higher order polynomials.  It is evident that higher order terms in
$F(X)$ can lead to higher order terms in $G(X)$.  So indeed one may
not be able to model $F(X)$ and $G(X)$ independently, provided the
Gaussian distribution for transverse fluctuations is assumed. \\ Let
us now consider longitudinal component of pressure gradient increment
${\cal P}_{x,u}=\langle \Delta \partial_x p|U,V,r\rangle$, which has a
major role in determining the main dynamical aspect of the inverse
cascade, i.e. the non equilibrium energy flux.  In contrast to the
transversal case, for the longitudinal case both $\langle \Delta
\partial_x p|U,r \rangle=\int_{-\infty}^{+\infty}{\cal P}_{x,u}
P(V|U,r) dV$ and $\langle \Delta \partial_x p|V,r
\rangle=\int_{-\infty}^{+\infty}{\cal P}_{x,u} P(U|V,r) dU$ are
non-trivial.  However, the resulting longitudinal PDF equation
involves only the $\langle \Delta \partial_x p|U,r\rangle$ and the
velocity conditional average $\langle V^2|U,r\rangle$, as one can
verify by integrating (\ref{pdf-joint}) over $V$.  Therefore, only the
knowledge of these two conditional averages is sufficient to close the
longitudinal PDF equation. Again the existence of the velocity
conditional
average indicates the importance of correlation effects in pure
longitudinal statistics.  The very existence of a non-equilibrium flux
implies that $P(U,r)\!=\!P(-\!U,-r)$, hence the PDF equation would
preserve the same invariance, i.e. $\langle
V^2|-\!U,-r\rangle\!=\!\langle V^2|U,r\rangle$ and $\langle \Delta
\partial_x p|-U,-r\rangle=-\langle \Delta\partial_x p|U,r\rangle$.
Scaling invariance of the PDF equation implies scaling invariance of
${\cal P}_{x,u}$ and $\langle V^2|U,r\rangle$.  Analogous to the
transversal case, we assume a local scale-invariant expansion for
$\langle \Delta\partial_x p|U,r\rangle$ and $\langle V^2|U,r\rangle$,
and we seek for a low order closure in terms of $Y={U}/{(\varepsilon
r)^{1/3}}$.  So defining $\langle \Delta\partial_x p|U,r\rangle
=[(\varepsilon r)^{2/3}/r] H(Y)$ and $\langle
V^2|U,r\rangle=(\varepsilon r)^{2/3} M(Y)$, we propose the following
expansion
\begin{eqnarray} 
H(Y)&=&E\left(Y^2-\frac{3}{5}M(Y)-\frac{6}{5A_{2,0}}Y\right)\,,
\label{H-eq} \\                           
M(Y)&=&A_{2,0}\left(\frac{5}{3}+\frac{Y}{2A_{2,0}^2}\right)\label{M-eq}\,.
\end{eqnarray}  
The coefficients of the three terms in the conditional pressure 
gradient are constrained by homogeneity, isotropy and incompressibility
(i.e. $\overline{YH(Y)}=0$ and $\overline{H(Y)}=0$). 
We observe that having reduced the expansion of $M(Y)$ at the first order,
the only new coefficient is the constant $E$. In Fig.~2 we
show the numerical evaluation of $H(Y)$ and $M(Y)$.
From the figure a low order expansion in terms of $Y$ can be inferred
for both these objects. However, concerning $M(Y)$ the result is
hardly distinguishable from an almost constant value.  From a best fit
we found $E=-0.39$ with an error bar around 20$\%$.  
If the longitudinal fluctuations were purely Gaussian
then these models might be considered as a better approximation for 
$H(Y)$ and $M(Y)$.  
However, the longitudinal statistics is just
nearly Gaussian, indeed the non zero flux implies a non-zero skewness
and to the non zero odd order structure functions
$S_{2n+1,0}(r)=A_{2n+1,0}(\varepsilon r)^{{(2n+1)}/{3}}$.
Furthermore a very important observation in \cite{BCV00} indicates
that the hyper-skewness of higher orders, i.e.
$S_{2n+1,0}/S_{2,0}^{(2n+1) / 2}$, increases with order and can not be
considered as a small parameter. So the expectation from any kind of
modeling for $H(Y)$ and $M(Y)$ is taking care of these fine details
of the distribution. It seems improbable to have access to these fine
details within a one parameter low order closure or other low order
models. As a quantitative check one can plug the low order expansion
in the longitudinal PDF equation. 
Then it is straightforward to obtain
the following prediction 
\bea
&&A_{2n+1,0}=\frac{2n}{2n(E+\frac{1}{3})+\frac{4}{3}} \Big( \Big[(2n-1)(2n-3)!!
\nonumber\\
&&+\frac{(3E+1)(2n-1)!!}{2}\Big]A_{2,0}^{n-1}+A_{2,0}
\Big(E+\frac{5}{3}\Big)A_{2n-1,0} \Big)
\eea
%%%%%%%%%%%%%%%%%%%%%%%%
\begin{figure}
\vspace{-.4cm}
\centerline{\includegraphics[scale=0.57,draft=false]{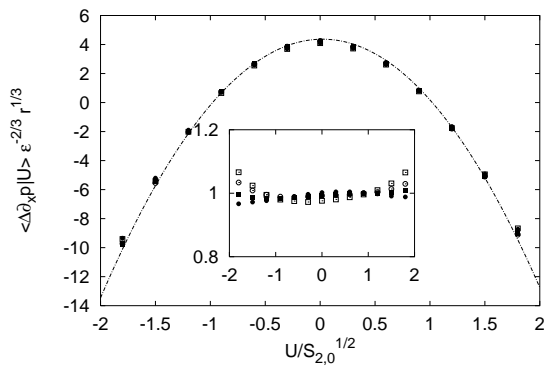}}
\narrowtext
\caption{$\langle \Delta \partial_x p|U,r \rangle$ computed at
$r=0.025$ (boxes) and $r=0.037$ (circles).  Empty
symbols refer to the Gaussian forcing and full ones to the restricted
in wavenumber one.  The full line is fitted with $E=-0.39$.  In the
inset we show $\langle V^2|U,r\rangle$ which up to two standard
deviations seems to be constant and at larger values fluctuates.}
\label{fig2}
%\vspace{-.4truecm}
\end{figure}
%%%%%%%%%%%%%%%%%%%%%%%%%%%%%%%%%%%%%%%%%%%%%%%%%  
Substituting the numerical value of $E$ we obtain for the
hyper-skewness $A_{5,0}/A_{2,0}^{5/2}\sim 0.449$ and
$A_{7,0}/A_{2,0}^{7/2}\sim 5.674$.  Comparing these numbers with the
corresponding numerically obtained values, $A_{5,0}/A_{2,0}^{5/2}\sim
0.25$ and $A_{7,0}/A_{2,0}^{7/2}\sim 1.55$, shows a large difference.
The fourth and sixth order hyper-flatnesses calculated from the closure
correspondingly are $A_{4,0}/A_{2,0}^2\sim3.29$ and
$A_{6,0}/A_{2,0}^3\sim20.03$. 
Comparing to the Gaussian values the deviations are getting bigger with
the order but still the errors are smaller in the even part with respect
to the odd part of the statistics.  
This is an important indication that, one has to consider higher
order expansions in order to be consistent with higher order statistics. 
Therefore, in spite of the fairly good compatibility between the low
order closure for $H(Y)$ and $M(Y)$ and their direct measurement in
two standard deviations the fine details of the distribution are not
recovered by them. This confirms the observation in \cite{BCV00} that
these fine details are buried in the very far tails of the
antisymmetric part of longitudinal PDF.\\
In conclusion, the dynamical role of the pressure gradient and
velocity conditional averages in establishing the velocity increments
statistics has been highlighted and numerically investigated.  The
transversal components of the velocity statistics has been found
Gaussian, in agreement with previous numerical and experimental
observations. Low order expansions for the transversal conditional
pressure gradient and $\langle U|V \rangle$, which have been proposed
(in a closely related approach) in the context of a self-consistent
closure \cite{Y01}, have been found in good agreement with the DNS
data up to two standard deviations. Furtherly, we proposed a
generalization of the expansion which is order by order consistent with
Gaussianity of the transverse statistics.  Concerning the
longitudinal statistics we found that the low order closure for the
conditional pressure gradient and $\langle V^2|U\rangle$, although in
fairly good compatibility with DNS data, is inconsistent with the fine
details of the longitudinal PDF, which bear the information of the
antisymmetric PDF tail. This indicates that unlike the transverse
statistics a complete description of the longitudinal statistics calls
for higher order terms in the expansions \cite{BCD01}. It is worth
emphasizing that these modelings are not just a naive fitting: the
free parameters are fixed {\it via} realisability conditions in the
PDF equations and have been tested numerically.  Let us finally remark
that the importance of the conditional averages goes far beyond the
assessment of closures for two-dimensional turbulence: the important
message is that any theoretical approach to pure longitudinal
(transversal) velocity statistics cannot disregard the reciprocal
dependence between longitudinal and transversal components. We
consider the investigation of such objects also in three dimensional
turbulence a necessary step.

We acknowledge discussions with A.~Celani, G.~Falkovich, T.~Gotoh,
R.J.~Hill, M.R.~Rahimitabar and V.~Yakhot.  G.B. has been partially
supported by the EU under the Grant ERB FMR XCT 98-0175
``Intermittency in Turbulent Systems'', and M.C.  by the the EU under
the Grant HPRN-CT-2000-00162) ``Non Ideal Turbulence''.
J.D. acknowledges the partial support by Deutsche
Forschungsgemeinschaft (DFG).  We also acknowledge the allocation of
computer resources from INFM ``Progetto Calcolo Parallelo''.
%%%%%%%%%%%%%%%%%%%%%%%%%%%%%%%%%%%%%%%%%%%%%%%%%%%%%%%%%%%%%

%%%%%%%%%%%%%%%%%%%%%%%%
%\begin{figure} 
%\vspace{-.3truecm}
%\centerline{\includegraphics[scale=0.60,draft=false]{fig1a.eps}}
%\centerline{\includegraphics[scale=0.60,draft=false]{fig1b.eps}}
%\vspace{0.4truecm}
%\narrowtext
%\vspace{-.2truecm}
%\caption{(a) $\langle U|V,r\rangle$ and (b) $\langle \Delta \partial_y
%p|V,r \rangle$ computed at $r= 0.025$ (boxes) and $r= 0.037$ 
%(circles).  Empty symbols refer to the Gaussian forcing and
%full ones to the restricted in wavenumber one.  The full lines
%represent predictions (\ref{pred1}) and (\ref{pred2}) with
%$A_{2,0}=11$. }
%\label{fig1}
%\vspace{-.3truecm}
%\end{figure}
%%%%%%%%%%%%%%%%%%%%%%%%%%%%%%%%%%%%%%%%%%%%%%%%%  
%\begin{figure}
%\vspace{-.4cm}
%\centerline{\includegraphics[scale=0.57,draft=false]{fig2.eps}}
%\narrowtext
%\caption{$\langle \Delta \partial_x p|U,r \rangle$ computed at
%$r=0.025$ (boxes) and $r=0.037$ (circles).  Empty
%symbols refer to the Gaussian forcing and full ones to the restricted
%in wavenumber one.  The full line is fitted with $E=-0.39$.  In the
%inset we show $\langle V^2|U,r\rangle$ which up to two standard
%deviations seems to be constant and at larger values fluctuates.}
%\label{fig2}
%\vspace{-.4truecm}
%\end{figure}
%%%%%%%%%%%%%%%%%%%%%%%%%%%%%%%%%%%%%%%%%%%%%%%%%  

\end{multicols}

\begin{thebibliography}{99}
\bibitem{nota} For $P(V,r)$, the quantity of
interest is the conditional average $\langle \Delta \partial_y p |
V,r\rangle \!=\! \int {\cal P}_{y,v} P(U|V,r) dU$. Due to rotational
symmetry one has parity invariance, i.e. $P(V,r)\!=\!P(-V,r)$, which
implies the invariance of $\langle \Delta \partial_y p|V,r\rangle$ for
$V\!\!\to\!\!  -V$. Taking into account that $\langle
U|V,r\rangle\!=\!\langle U|\!-V,r\rangle$, $\langle \Delta \partial_y
p|V,r\rangle$ would be an odd function of $V$.  Moreover, $\overline
{\langle \Delta \partial_y p|V,r\rangle}\!\equiv \!\int \langle \Delta
\partial_y p|V,r\rangle P(V) dV\!\!=\!\!0$.
\bibitem{nota2}In order to guarantee the correct
dimensionality, the Kolmogorov scaling $V\!\sim\!  r^{1/3}$ has been
imposed in (\ref{pre-eq1}), i.e. we implicitly assumed the absence of
intermittency, which should be consistent with the outcomes of the
theory.
\end{thebibliography}
\end{document}